\documentclass{elsart}

\usepackage{amsmath}
\usepackage{amsbsy}
\usepackage{amssymb}
\usepackage{latexsym}
\usepackage[dvips]{color,graphicx}

\begin{document}

\begin{frontmatter}
\title{Numerical simulations of strongly correlated
fermions confined in 1D optical lattices}
\author{Marcos Rigol} and
\author{Alejandro Muramatsu}

\address{Institut f\"ur Theoretische Physik III, 
Universit\"at Stuttgart, Pfaffenwaldring 57, D-70550 Stuttgart, Germany.}

\begin{abstract}
On the basis of quantum Monte Carlo (QMC) simulations we study 
the formation of Mott domains in the one-dimensional Hubbard model 
with an additional confining potential. We find evidences of 
quantum critical behavior at the boundaries of the Mott-insulating 
regions. A local compressibility defined to characterize the local 
phases exhibits a non-trivial critical exponent on entering the 
Mott-insulating domains. Both the local compressibility and the variance 
of the local density show universality with respect to the confining 
potential. We also study the momentum distribution function of the 
trapped system, and determine its phase diagram.
\end{abstract}
\begin{keyword}
Optical lattices \sep metal--Mott-insulator transition \sep
Degenerated Fermi gases \sep Strongly correlated systems  
\PACS 03.75.Ss \sep 05.30.Fk \sep 71.30.+h
\end{keyword}
\date{}
\end{frontmatter}

\section{Introduction}

\vspace{-0.3cm}
 
The study of ultracold quantum gases has become in the last decade 
a field of intense experimental and theoretical research 
\cite{dalfovo99,leggett01,pethick02,pitaevskii03}. Bose-Einstein
condensation (BEC) was the main motivation starting the intensive analysis 
of such systems. It was observed for the first time in a series of 
experiments on dilute vapors of alkali atoms cooled down to extremely low 
temperatures (fractions of microkelvins) 
\cite{anderson95,bradley95,davis95}.

Recently, new features that allow to go beyond the weakly interacting 
regime and access strongly correlated limits have been added to the 
experiments. Particularly important, due to its relevance for 
condensed matter physics, has been the introduction of 
optical lattices. They allowed the study of the 
superfluid--Mott-insulator phase transition in three-dimensional 
\cite{greiner02}, and one-dimensional (1D) \cite{stoferle04} systems. 
The presence of the optical lattice and the fact that 
the particles interact only via contact interaction, lead in a natural 
way to the Hubbard model as a paradigm for these systems. A theoretical 
work proposing such experiments \cite{jaksch98}, and quantum Monte Carlo 
(QMC) simulations \cite{batrouni02,kashurnikov02,wessel04} 
have examined these systems in detail. It has been found that  
incompressible Mott insulating phases appear for wide ranges of fillings, 
and always coexist with compressible phases, so that local order parameters 
\cite{batrouni02,wessel04} have to be defined to characterize the system.

In the fermionic case, the metal--Mott-insulator transition (MMIT) 
\cite{rigol03_1,rigol03_2} has not been observed yet. 
However, recent experiments succeeded in loading single species ultracold 
fermionic gases on an optical lattice \cite{modugno03,ott04}, allowing the 
realization of an ideal Fermi gas on a lattice. Progress in this field, 
loading more than one component fermions and reducing the occupation per 
lattice site, could lead to the realization of the metal--Mott-insulator 
transition on optical lattices.

Motivated by this expectation we studied, using QMC simulations, 
the ground state of the one dimensional (1D) fermionic Hubbard model
with a harmonic trap \cite{rigol03_1,rigol03_2}. In the present work 
we review those references adding some new results and improving others.
As in the bosonic case \cite{batrouni02,kashurnikov02,wessel04}, 
we find that Mott domains appear over a continuous range of fillings and 
always coexist with compressible phases. Therefore, we define a local order 
parameter, that we call local compressibility, in order to characterize the 
local phases present in the system. By means of this local compressibility, 
we analyze in detail the interface between the metallic and insulating 
region finding that critical behavior sets in, revealing a new critical 
exponent. Furthermore, the behavior of the local compressibility and the 
variance of the density are found to be universal in this case 
independently of the confining potential and the strength of the 
interaction. Hence, universality appears as usual in critical phenomena. 
The momentum distribution function, a quantity usually accessed in the 
experiments, is studied in detail. 
The results obtained show that due to the inhomogeneous character of these 
systems, this quantity does not exhibit a clear signature of the formation of 
Mott domains. Finally, we obtain the phase diagram for fermions confined in 
1D harmonic traps. It allows to compare systems with different fillings 
and curvatures of the confining potential, so that it could help to 
understand future experimental results.

The exposition is organized as follows. In the next section, we give 
a short introduction to the Hubbard model on periodic systems. In 
Sec.\ \ref{LocalC}, we study local quantities within the Hubbard model 
with an additional harmonic trap. In particular, we define a local 
compressibility that acts as a local order parameter to characterize the 
Mott-insulating phases. In Sec.\ \ref{PRL}, we study in detail 
the region where the transition between the metallic and the Mott-insulating 
phases occurs. The momentum distribution function is analyzed in 
Sec.\ \ref{MDF}. In Sec.\ \ref{FD}, we discuss the phase diagram for these 
systems. Finally, the conclusions are given in Sec.\ \ref{Conc}.

\section{The Hubbard model}

The Hubbard model in a periodic lattice \cite{anderson59,hubbard63} 
has been intensively studied in condensed matter physics as a prototype 
for the theoretical understanding of the MMIT. This model considers 
electrons in a single band, and its Hamiltonian can be written as
\begin{equation}
H  =  -t \sum_{i,\sigma} \left( c^\dagger_{i\sigma} c^{}_{i+1
\sigma} + \textrm{H.c.} \right) + U \sum_i n_{i \uparrow} n_{i \downarrow}
\label{HubbH}
\end{equation}
where $c^\dagger_{i\sigma}$, $c_{i\sigma}$ are creation and 
annihilation operators, respectively, for electrons with spin $\sigma$ 
at site $i$, and $n_{i \sigma} = c^\dagger_{i\sigma} c^{}_{i\sigma}$ 
is the particle number operator. Electrons are considered to hop only 
between nearest neighbors with a hopping amplitude $t$, and the on-site 
interaction parameter is denoted by $U$ ($U>0$). 

\begin{figure}[h]
\begin{center}
\includegraphics[width=0.6\textwidth,height=0.44\textwidth]
{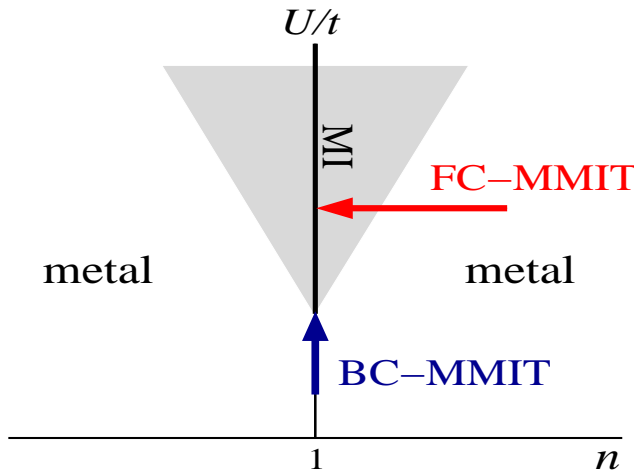}
\end{center}
\caption
{General schematic phase diagram for the metal--Mott-insulator 
transition obtained from the Hubbard model. The shaded area represents 
a metallic region that is under the strong influence of the MMIT. 
The routes for the MMIT (see text) has been signaled with arrows.}
\label{HubbPD}
\end{figure}
It is well known that the Hubbard model [Eq.\ (\ref{HubbH})] displays 
MMIT phase transitions at half filling and at finite values of the on-site
repulsive interaction $U$ \cite{imada98}. (In the case of perfect
nesting\footnote{Nesting refers to the existence of parallel sections 
in the Fermi surface. This means that excitations of vanishing energy 
are possible at finite momentum. Perfect nesting appears for periodic 
systems in 1D at any filling since in 1D the Fermi surface consists only of 
two points. In addition, it appears for hypercubic lattices at half filling 
since in those cases the Fermi surface is a hypercube.} the transition occurs 
at $U=0$.) The schematic phase diagram is shown in Fig.\ \ref{HubbPD}, in the 
plane $U/t$ vs $n$. The two routes for the MMIT are the filling-controlled 
MMIT (FC-MMIT), and the bandwidth-controlled MMIT (BC-MMIT) \cite{imada98}. 
Metallic regions very close to the 
Mott-insulating phase [shaded area in Fig.\ \ref{HubbPD}] are under the 
strong influence of the MMIT, and in general exhibit anomalous 
features like mass enhancement and carrier number reduction \cite{imada98}. 
In 1D systems, the metallic phase (Luttinger liquid) is characterized by 
gapless excitations and by a $2k_F$ (where $k_F$ is the Fermi momentum) 
modulation of the spin-spin correlations. On the other side, in the 
Mott-insulating phase ($n=1$) the system exhibits a charge gap, and 
quasi-long range antiferromagnetic correlations. At $n=0,2$ the system 
is in a trivial band insulating state. 

\section{The Hubbard model in a harmonic trap \label{LocalC}}

Ultracold fermionic atoms on optical lattices represent nowadays the most 
direct experimental realization of the Hubbard model, since atoms interact 
only via a contact potential. In addition, in the experiments all 
parameters can be controlled with an unprecedented precision. Since the 
particles are trapped by a confined potential, it is possible to drive 
the MMIT by changing the total filling of the trap, the on-site
repulsive interaction, or varying the curvature of the confining 
potential. The Hamiltonian in this case can be written as
\begin{equation}
H = -t \sum_{i,\sigma} \left( c^\dagger_{i\sigma} c^{}_{i+1
\sigma} + \textrm{H.c.} \right) + U \sum_i n_{i \uparrow} n_{i \downarrow}
+ V_2 \sum_{i \sigma} x_i^2\ n_{i \sigma}, \label{Hubb}
\end{equation}
where $V_2$ is the curvature of the harmonic confining potential, 
and $x_i$ measures the position of the site $i$ ($x_i=ia$ with $a$ 
the lattice constant). The number of lattice sites is $N$, and is selected 
so that all the fermions are confined in the trap. We denote the total 
number of fermions as $N_f$ and consider equal number of 
fermions with spins up and down ($N_{f\uparrow}=N_{f\downarrow}=N_f/2$). 
In our simulations, we used the zero-temperature
projector QMC method described in detail in Refs.\ 
\cite{loh92,muramatsu99,assaad02}.

\begin{figure}[h]
\begin{center}
\includegraphics[width=0.63\textwidth,height=0.4\textwidth]
{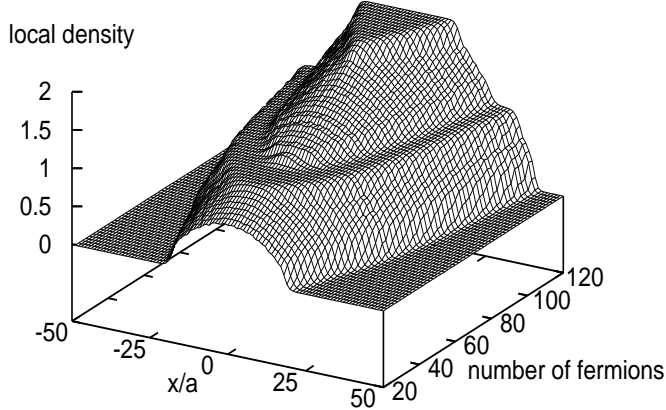}
\end{center}
\caption{Evolution of the local density in a parabolic confining 
potential as a function of the position in the trap and increasing total 
number of fermions. The parameters involved are $N=100$, $U=6t$ and 
$V_2a^2=0.006t$. The positions are measured in units of the 
lattice constant $a$.}
\label{perfil3D}
\end{figure}
Results for the evolution of the local density ($n_{i}=\langle
n_{i\uparrow }+n_{i\downarrow}  \rangle$), as a function of the 
total number of the confined particles, are shown in Fig.\ \ref{perfil3D}.
For the lowest fillings, so that $n < 1$ at every site, the density shows 
a profile with the shape of an inverted parabola, similar to that 
obtained in the non-interacting case \cite{vignolo00}, and hence, 
such a situation should correspond to a metallic phase.
Increasing the number of fermions a plateau with
$n=1$ appears in the middle of the trap, surrounded by a region with
$n<1$ (metallic). Since in the periodic case, a Mott insulator appears 
at $n=1$, it is natural to identify the plateau with such a phase. 
The Mott-insulating domain in the center of the trap
increases its size when more particles are added, but at a certain
filling this becomes energetically unfavorable and 
a new metallic phase with $n > 1$ starts to develop in the center 
of the system. Upon adding more fermions, this new 
metallic phase widens spatially and the Mott-insulating domains surrounding it 
are pushed to the borders. Depending on the on-site repulsion strength,
they can disappear and a complete metallic phase can appear in the
system. Finally, a ``band insulator'' (i.e., $n=2$) forms in the middle of 
the trap for the highest fillings. Due to
the full occupancy of the sites, it will widen spatially and pushes the
other phases present in the system to the edges of the trap when
more fermions are added.

\begin{figure}[h]
\begin{center}
\includegraphics[width=1.0\textwidth,height=0.8\textwidth]
{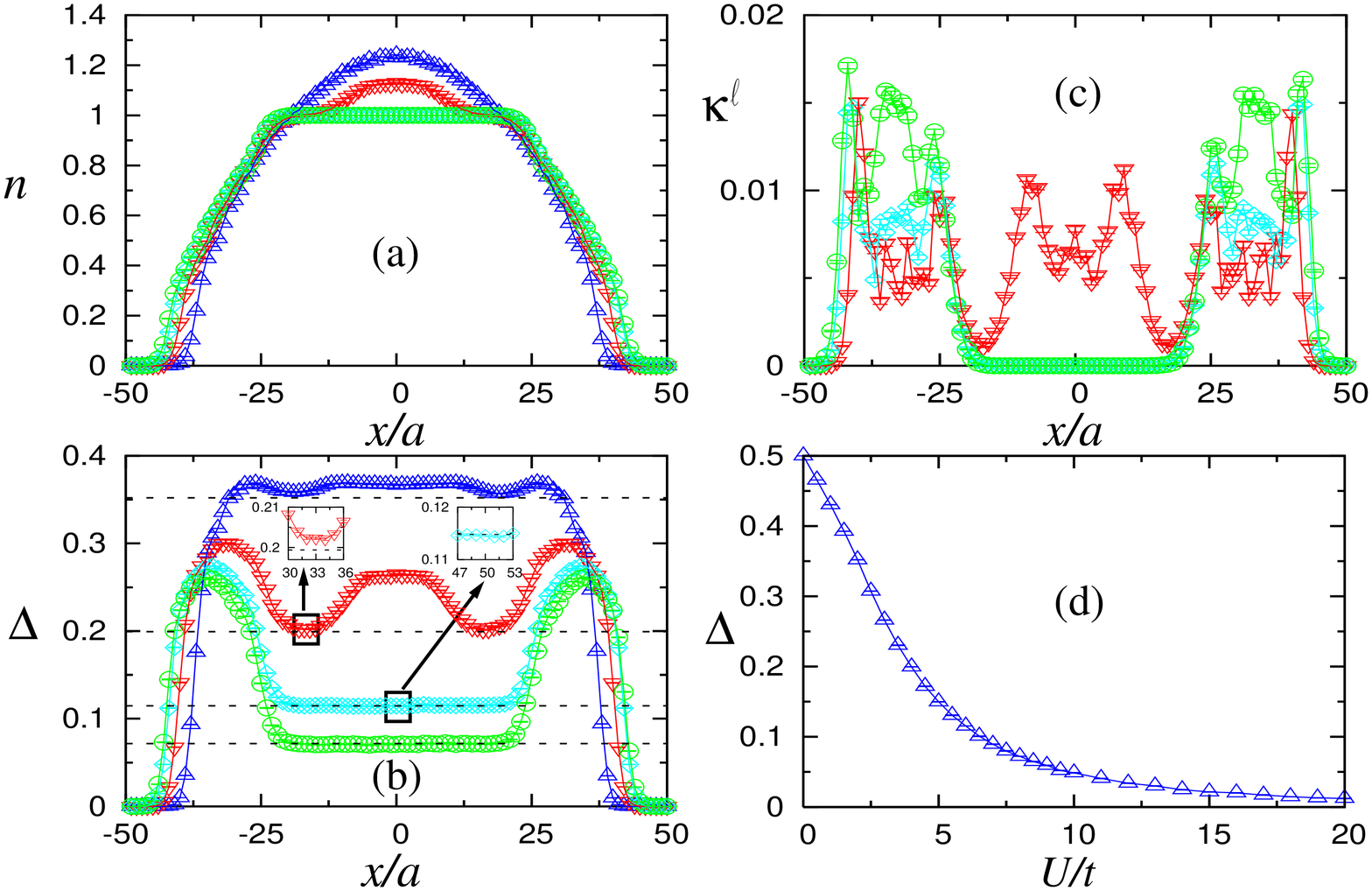}
\end{center}
\caption{Profiles for a trap with $V_2a^2=0.0025t$ and $N_f=70$, 
the on-site repulsions are $U/t=$2 (\textcolor{blue}{$\triangle$}), 
4 (\textcolor{red}{$\bigtriangledown$}), 
6 (\textcolor{cyan}{\large $\Diamond$}), 
and 8 (\textcolor{green}{$\bigcirc$}). (a) Local density, 
(b) variance of the local density, (c) local compressibility 
$\kappa^\ell$ as defined in Eq.\ (\ref{localc}), (d) variance of the 
density in the Mott-insulating state (when $U/t>0$) vs $U/t$. 
The dashed lines in (b) are the values of the variance in the $n=1$ 
periodic system for $U/t=2, \ 4,\ 6,\ 8$ (from top to bottom).}
\label{perfdeltcomp100}
\end{figure}

Alternatively to the case shown in Fig.\ \ref{perfil3D}, where the 
transitions between the phases were driven by an increment of the 
filling in the trap, Mott-insulating regions can be obtained 
by increasing the ratio $U/t$. Fig.\ \ref{perfdeltcomp100}(a) shows 
the evolution of the density profiles in a trapped system 
when this ratio is increased from $U/t=2$ to $U/t=8$. 
It can be seen that for small values of $U/t$ ($U/t=2$) there is only 
a metallic phase present in the
trap. As the value of $U/t$ ($U/t=4$) is increased, a Mott-insulating 
phase tries to develop at $n=1$ while a metallic phase with $n>1$ is present
in the center of the system. As the on-site repulsion is increased even 
further ($U/t=6,\ 8$), a Mott-insulating domain appears in the middle of the 
trap suppressing the metallic phase that was present there. 
In Fig.\ \ref{perfdeltcomp100}(b) we show the variance of the density for 
the profiles in Fig.\ \ref{perfdeltcomp100}(a) (from top to bottom, the 
values presented are for $U/t=2, \ 4, \ 6,\ 8$). As expected, the 
variance decreases in both the metallic and Mott-insulating phases when the 
on-site repulsion is increased. When the Mott-insulating plateau is formed in 
the density profile, a plateau with constant variance appears in the variance
profile with a value that will vanish only in the limit
$U/t \rightarrow \infty$. The dependence of the variance in the Mott 
insulator vs $U/t$ is depicted in Fig.\ \ref{perfdeltcomp100}(d) up to
$U/t=20$ (five times the band-width). In this figure it is possible to see 
that after a fast decrease up to around $U/t=8$, the variance reduces slowly 
[proportionally to $(U/t)^{-2}$] when increasing U. As shown in 
Fig.\ \ref{perfdeltcomp100}(b), whenever a Mott-insulating domain is 
formed in the trap, the value of the variance on it is exactly the same 
as the one for the Mott-insulating phase in the periodic system for the 
same value of $U/t$ (horizontal dashed lines). 
This would support the validity of the commonly used local density 
(Thomas-Fermi) approximation \cite{butts97}. However, the insets in 
Fig.\ \ref{perfdeltcomp100}(b), show that this is not necessarily the 
case, since for $U/t=4$, the value of the variance in the Mott-insulating 
phase of the periodic system is still not reached in the trap, although the 
density reaches the value $n = 1$. Therefore, in contrast to the periodic
case, a Mott-insulating region is not only determined by the filling.  
In the cases of $U/t=6$ [inset in Fig.\ \ref{perfdeltcomp100}(b) for a 
closer look] and $U/t=8$, the value of the variance in the periodic 
system is reached and then one can say that Mott-insulating phases are formed 
there.

Although the variance gives a first indication for the formation of
a local Mott insulator, an ambiguity is still present, since
there are metallic regions with densities very close to $n=0$ and 
$n=2$, where the variance can have even smaller values than in 
the Mott-insulating phases. Therefore, an unambiguous quantity is still
needed to characterize the Mott-insulating regions.
We proposed a new local compressibility as a local-order parameter
to characterize the Mott-insulating regions\cite{rigol03_1,rigol03_2}, 
that is defined as
\begin{equation}
\label{localc} \kappa_i^\ell = \sum_{\mid j \mid \leq \, \ell (U)}
\chi_{i,i+j} \ ,
\end{equation}
where
\begin{equation}
\chi _{i,j}=\left\langle n_{i}n_{j}\right\rangle -\left\langle
n_{i} \right\rangle \left\langle n_{j}\right\rangle
\end{equation}
is the density-density correlation function and $\ell (U) \simeq b
\, \xi (U)$, with $\xi (U)$ the correlation length of $\chi_{i,j}$
in the unconfined system at half-filling for the given value of $U$. 
As a consequence of the charge gap opened in the Mott-insulating 
phase at half filling in the periodic system, the
density-density correlations decay exponentially [$\chi_{\left( x
\right)} \propto \exp^{-\frac{x}{\xi\left( U\right)}}$] enabling 
$\xi (U)$ to be determined. The factor $b$ is chosen within a 
range where $\kappa^\ell$ becomes qualitatively insensitive to its 
precise value ($b\sim 10$) \cite{rigol03_2}. Physically, the local 
compressibility defined here gives a measure of the change in the 
local density due to a constant shift of the potential over a finite range 
but over distances larger than the correlation length in the unconfined 
system.
 
In Fig.\ \ref{perfdeltcomp100}(c), we show the profiles of the local 
compressibility for the same parameters as Figs.\
\ref{perfdeltcomp100}(a) and (b) (we did not include the profile of the 
local compressibility for $U=2t$ because for that value of $U$ we obtain 
that $\ell$ is bigger than the system size). In Fig.\ 
\ref{perfdeltcomp100}(c), it can be seen that the local compressibility 
only vanishes in the Mott-insulating domains. For $U=4t$, it can be seen
that in the region with $n \sim 1$ the local compressibility,
although small, does not vanish. This is compatible with the fact that 
the variance is not equal to the value in the periodic 
system there, so that although there is a shoulder in the density profile, 
this region is not a Mott insulator. Therefore, the local compressibility 
defined here serves as a genuine local order parameter to
characterize the insulating regions that always coexist with
metallic phases.

\section{Local quantum criticality and universality \label{PRL}}

In the previous section we have characterized the local Mott-insulating 
and metallic phases quantitatively. We study in this section
the regions where the system goes from one phase to another. 
Criticality can arise, despite the microscopic spatial size 
of the transition region\footnote{Recent experiments leading to a MMIT 
\cite{greiner02,stoferle04} considered a systems with linear dimensions 
$\sim 100 a$, i.e.\ still in a microscopic range.}, due to the extension 
in imaginary time that reaches the thermodynamic limit at $T=0$, 
very much like the case of the single impurity Kondo problem 
\cite{yuval70}, where long-range interactions in imaginary time appear 
for the local degree of freedom as a result of the interaction with the 
rest of the system. An intriguing future question, for both theory
and experiment, will be the role of spatial dimension
in the critical behavior of systems in the thermodynamic limit. 
In the periodic case, the FC-MMIT in 1D and 2D belong 
to different universality classes, with dynamical exponents 
$z=2$ (1D) and $z=4$ (2D) \cite{imada98}. This route 
to the transition in the periodic case, is the one relevant for the 
trapped systems where the density changes on entering in the local 
Mott insulator, and could lead to different local quantum critical 
behavior between traps with different dimensionalities. In the 
bosonic case the FC-MMIT exhibits the same dynamical exponent ($z=2$)
in all dimensions \cite{fisher89,batrouni90}, which 
could also lead to a different local quantum critical 
behavior for trapped bosons as compared with trapped fermions. 

\begin{figure}[h]
\begin{center}
\includegraphics[width=0.66\textwidth,height=0.43\textwidth]
{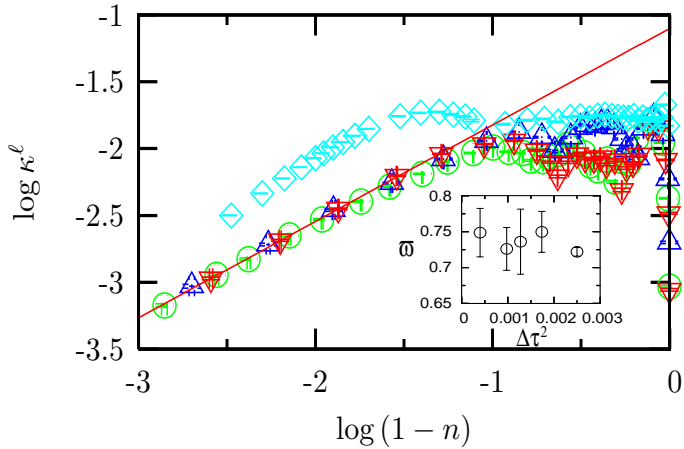}
\end{center} 
\caption{The local compressibility $\kappa^{\ell}$ vs
$\delta=1-n$ at $\delta \rightarrow 0$ for
(\textcolor{blue}{$\triangle$}) $N_f = 70$, $U=8t$ and $V_2a^2=0.0025t$; 
(\textcolor{red}{$\bigtriangledown$}) $N_f = 70$, $U=6t$ and $V_2a^2=0.0025t$; 
(\textcolor{green}{$\bigcirc$}) $N_f=72$, $U=6 t$ and a quartic potential 
with $V_4a^4=1.0\times 10^{-6}t$; 
(\textcolor{cyan}{$\Diamond$}) unconfined periodic system with $U=6t$.
The straight line displays a power-law behavior $\varpi=0.72$.
Inset: Dependence of the critical exponent $\varpi$ on $\Delta \tau^2$.}
\label{kappavsn}
\end{figure}
\begin{figure}[h]
\begin{center}
\includegraphics[width=0.57\textwidth,height=0.39\textwidth]
{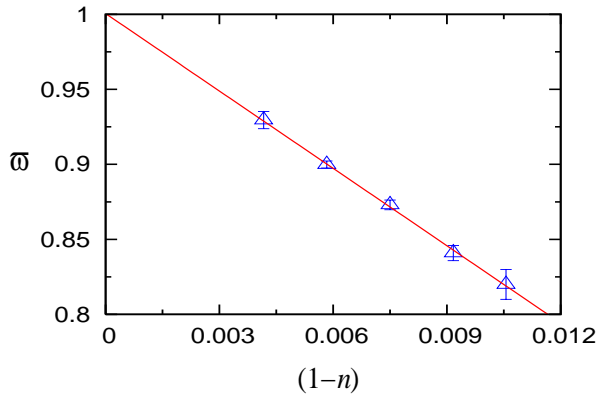}
\end{center} 
\caption{Exponent $\varpi$ of the local compressibility for the 
periodic system, obtained between contiguous points in 
Fig.\ \ref{kappavsn}, vs $(1-n)$. The extrapolation leads 
to $\varpi=1$ for $n\rightarrow 1$.}
\label{Expcompvsn}
\end{figure}

 
Figure \ref{kappavsn} shows the local compressibility vs 
$\delta=1-n$ for $\delta \rightarrow 0$ in a double logarithmic plot. A
power law $\kappa^{\ell} \sim \delta^\varpi$ is obtained, with $\varpi <
1$, such that a divergence results in its derivative with respect to
$n$, showing that critical fluctuations are present in this region.
Since the QMC simulation is affected by systematic errors due to
discretization in imaginary time, it is important to consider the
limit $\Delta \tau \rightarrow 0$ in determining the critical
exponent. The inset in Fig.\ \ref{kappavsn} shows such an
extrapolation leading to $\varpi \simeq 0.68 - 0.78$.  At this point
we should remark that the presence of the harmonic potential allows the 
determination of the density dependence of various quantities with 
unprecedented detail on feasible system sizes as opposed to unconfined 
periodic systems, where systems with $10^3 - 10^4$ sites would be necessary 
to allow for similar variations in density. In addition to the power law
behavior, Fig.\ \ref{kappavsn} shows that for $\delta
\rightarrow 0$, the local compressibility of systems with a harmonic
potential but different strengths of the interaction or even with a
quartic confining potential, collapse on the same curve. 
Hence, universal behavior as expected for critical
phenomena is observed also in this case.  This fact is particularly
important with regard to experiments, since it implies that the
observation of criticality should be possible for realistic confining
potentials, and not only restricted to perfect harmonic ones, as usually
used in theoretical calculations. However, Fig.\ \ref{kappavsn} shows 
also that the unconfined case departs from all the others. Up to the 
largest systems we simulated (600 sites), we observe an increasing slope 
rather than the power law of the confined systems. Actually, we observe 
that the exponent of the power law obtained between contiguous points 
in Fig.\ \ref{kappavsn} for the periodic case extrapolates to 
$\varpi=1$, as it is shown in Fig.\ \ref{Expcompvsn}. 

\begin{figure}[h]
\begin{center}
\includegraphics[width=0.77\textwidth,height=0.55\textwidth]
{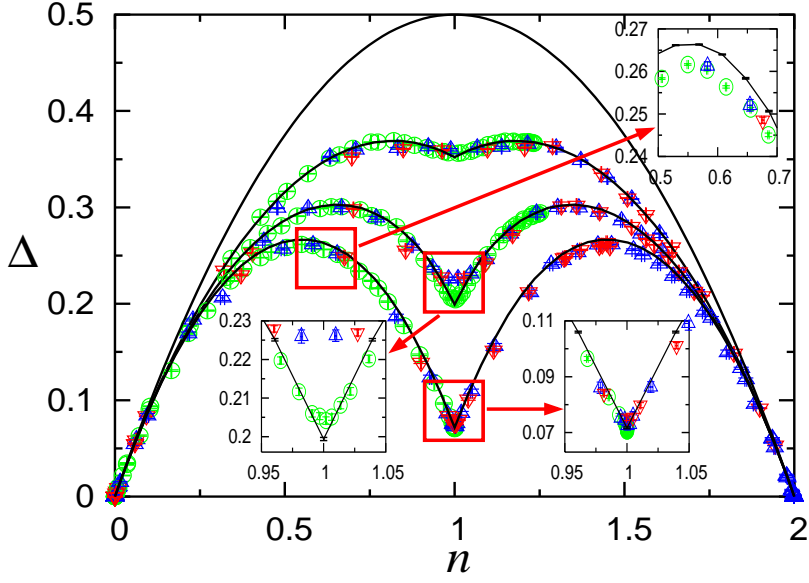}
\end{center} 
\caption{Variance $\Delta$ {\em vs.} $n$ for
(\textcolor{green}{$\bigcirc$}) harmonic potential $V_2a^2=0.0025 t$ 
with $N=100$; (\textcolor{blue}{$\triangle$}) quartic
potential $V_4a^4=5\times 10^{-7} t$ with $N=150$;
(\textcolor{red}{$\bigtriangledown$}) harmonic potential $V_2a^2=0.016 t$ + 
cubic $V_3a^3 = 1.6\times 10^{-4} t$ + quartic $V_4a^4 = 1.92\times 10^{-5} t$ 
with $N=50$; and (full line) 
unconfined periodic potential with $N=102$ sites. The curves correspond 
from top to bottom to $U/t = 0, 2, 4, 8$. For a discussion of the insets, 
see text.}
\label{DeltaVsnAll}
\end{figure}

\begin{figure}[h]
\begin{center}
\includegraphics[width=0.66\textwidth,height=0.43\textwidth]
{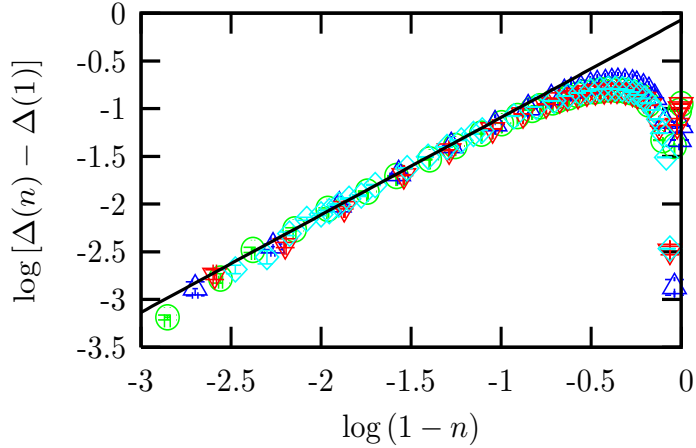}
\end{center} 
\caption{Variance of the density vs
$\delta=1-n$ at $\delta \rightarrow 0$ for
(\textcolor{blue}{$\triangle$}) $N_f = 70$, $U=8t$ and $V_2a^2=0.0025t$; 
(\textcolor{red}{$\bigtriangledown$}) $N_f = 70$, $U=6t$ and $V_2a^2=0.0025t$; 
(\textcolor{green}{$\bigcirc$}) $N_f=72$, $U=6 t$ and a quartic potential 
with $V_4a^4=1.0\times 10^{-6}t$; (\textcolor{cyan}{$\Diamond$}) unconfined 
periodic system with $U=6t$. The straight line displays linear behavior, 
i.e.\ its slope is equal to one.}
\label{VaroccupPaper}
\end{figure}

Having shown that the local compressibility displays universality on
approaching a Mott-insulating region, we consider the variance
$\Delta$ as a function of the density $n$ for various values of $U$
and different confining potentials. Figure \ref{DeltaVsnAll} shows
$\Delta$ vs $n$ for a variety of systems, where not only the
number of particles and the size of the system are changed, but also
different forms of the confining potential were used.  Here we
considered a harmonic potential, a quartic one, and a superposition of
a harmonic, a cubic and a quartic one, such that even reflection
symmetry across the center of the system is broken. It appears at
first glance that the data can only be distinguished by the strength
of the interaction $U$, showing that the variance is rather
insensitive to the form of the potential.  The different insets,
however, show that a close examination leads to the conclusion that
only near $n=1$ and only in the situations where at $n=1$ a
Mott insulator exists, universality sets in.
The inset for $n$ around 0.6 and $U=8t$, shows that the unconfined
system has different variance from the others albeit very close on a
raw scale. This difference is well beyond the error bars. Also the inset 
around $n=1$ and for $U=4t$, shows that systems that do not form a 
Mott-insulating phase in spite of reaching a density $n=1$, have a different 
variance from those having a Mott insulator. Only the case where all 
systems have a Mott-insulating phase at $n=1$ ($U=8t$), shows universal 
behavior independent of the potential, a universality that encompasses 
also the unconfined systems. 
For the unconfined system, the behavior 
of the variance can be examined with Bethe-{\em Ansatz} \cite{lieb68} 
in the limit $\delta \rightarrow 0$. In this limit and to leading order 
in $\delta$, the ground state energy is given by \cite{schadschneider91} 
$E_0 (\delta)/N - E_0 (\delta=0)/N \propto \delta$, such that the double 
occupancy, which can be obtained as the derivative of the ground-state 
energy with respect to $U$, will also converge as $\delta$ towards its 
value at half-filling. Such behavior is also obtained in our case as shown 
in Fig.\ \ref{VaroccupPaper} for the same parameters of Fig.\ \ref{kappavsn}.

\section{Momentum distribution function \label{MDF}}

In most experiments with quantum gases carried out so far,
the MDF, which is determined in time-of-flight measurements, 
played a central role. A prominent example is given by the study of the 
superfluid--Mott-insulator transition \cite{greiner02} in the bosonic case.
As shown below, we find that the MDF is not appropriate to characterize the 
phases of the system in the fermionic case, and does not show any clear 
signature of the MMIT.

\begin{figure}[h]
\begin{center}
\includegraphics[width=0.77\textwidth,height=0.66\textwidth]
{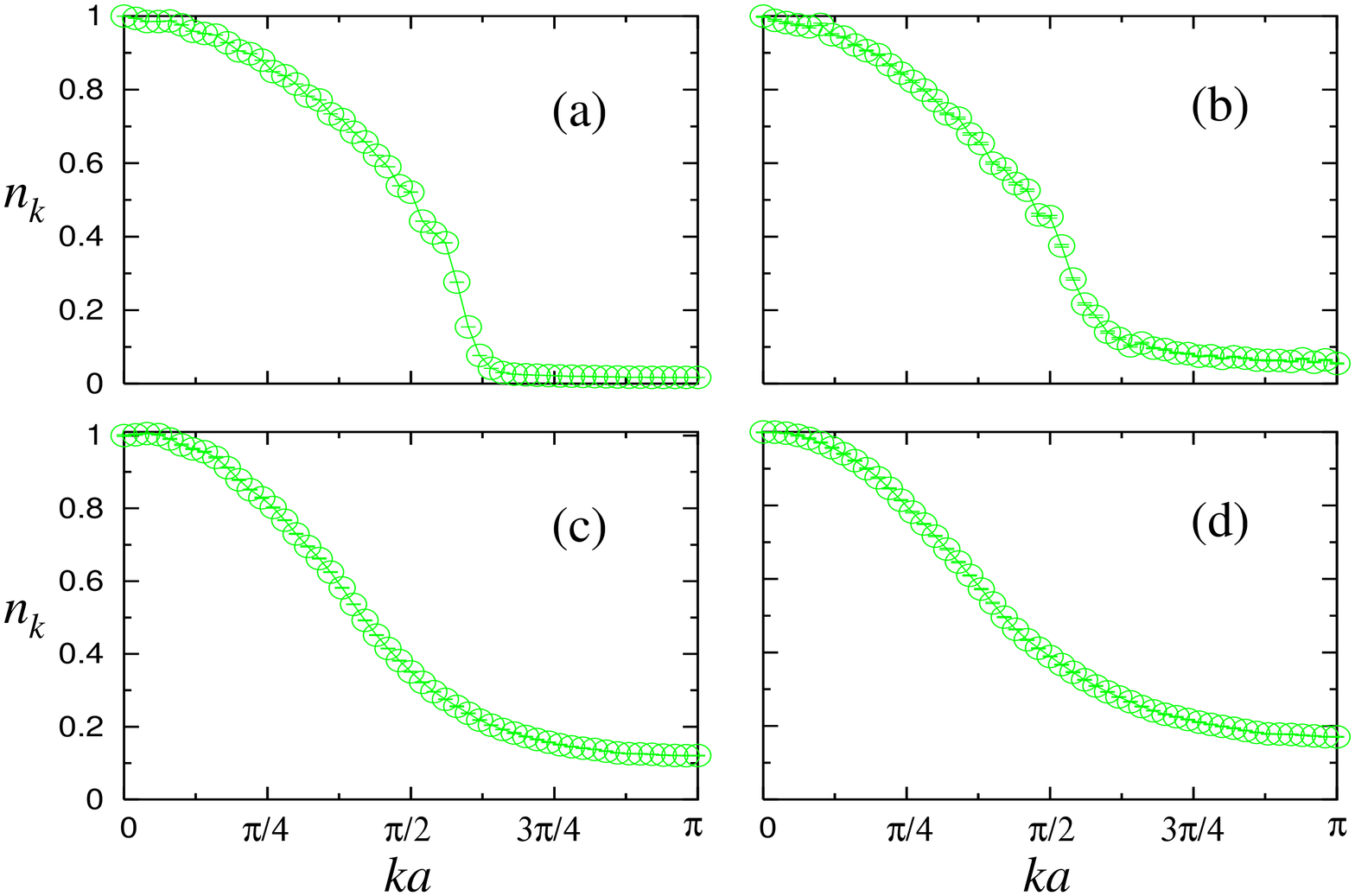}
\end{center} 
\caption{Normalized momentum distribution function for 
$U/t=2$ (a) 4 (b), 6 (c), 8 (d), 
and $N=100$, $N_f=70$, $V_2a^2=0.0025t$.}
\label{PerfilK100}
\end{figure}
In Fig.\ \ref{PerfilK100} we show the normalized momentum 
distribution function for the density profiles shown in 
Fig.\ \ref{perfdeltcomp100}. For the trapped systems, we always 
normalize the MDF to be unity at $k=0$. We first notice that $n_k$ 
for the pure metallic phase in the harmonic trap 
[Fig.\ \ref{PerfilK100}(a)] does not
display any sharp feature corresponding to a Fermi surface, in clear 
contrast to the periodic case. The lack of a sharp feature for the 
Fermi surface is independent of the presence of the interaction and is 
also independent of the size of the system. In the non-interacting case,
this can be easily understood: the spatial density and the
momentum distribution will have the same functional form because
the Hamiltonian is quadratic in both coordinate and momentum. When
the interaction is present, it could be expected that the formation of
local Mott-insulating domains generates a qualitatively and quantitatively
different behavior of the momentum distribution, like in the
periodic case where in the Mott-insulating phase the Fermi
surface disappears. In Fig.\ \ref{PerfilK100}(c), 
it can be seen that there is no qualitative change of the MDF 
when the Mott-insulating phase is present in the
middle of the trap. Quantitatively $n_k$ in this case is 
similar to the pure metallic cases Fig.\ \ref{PerfilK100}(a),(b). 
Quantitative changes in $n_k$ appear only when the on-site repulsion 
goes to the strong-coupling regime, but this is long after the 
Mott-insulating phase has appeared in the system.

At this point one might think that in order to study the MMIT using 
the MDF, it is necessary to avoid the 
inhomogeneous trapping potential and use instead a kind of magnetic box 
with infinitely high potential on the boundaries. However, in that case 
one of the most important achievements of the inhomogeneous system is 
lost, i.e., the possibility of creating Mott-insulating phases for a 
continuous range of fillings. In the perfect magnetic box, the 
Mott-insulating phase would only be possible at half filling, which would be 
extremely difficult (if possible at all) to adjust experimentally. 
The other possibility is to create traps 
which are almost homogenous in the middle and which have an appreciable 
trapping potential only close to the boundaries. This can be
studied theoretically by considering traps with higher powers of the
trapping potentials. As shown below, already non-interacting systems make 
clear that a sharp Fermi edge is missing in confined systems.

\begin{figure}[h]
\begin{center}
\includegraphics[width=0.96\textwidth,height=0.36\textwidth]
{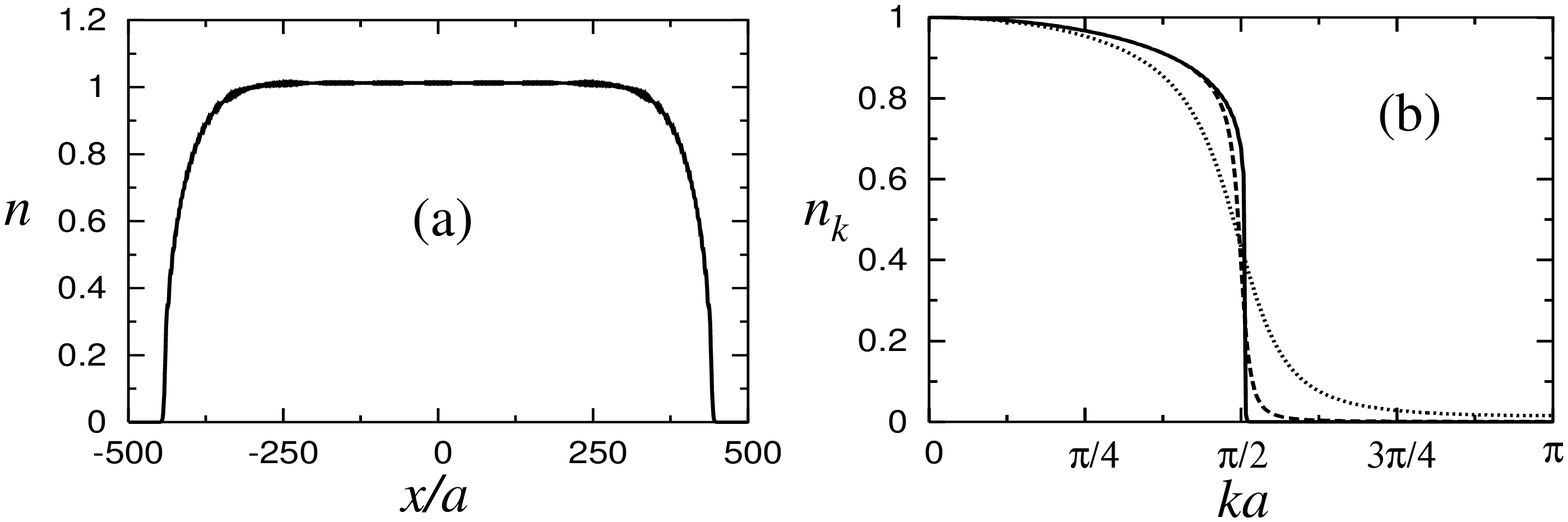}
\caption{Exact results for $N_f=840$ noninteracting trapped fermions in 
a lattice with 1000 sites and a confining potential 
$V_{10}a^{10}=7 \times 10^{-27}t$. Density profile (a) and the normalized 
momentum distribution function (b): the continuous line corresponds to 
(a), the dashed line is the result when an alternating potential 
$V_a=0.1t$ is superposed on the system, and the dotted line corresponds 
to $V_a=0.5t$.}
\label{PerfilK1000}
\end{center}
\end{figure}
In Fig.\ \ref{PerfilK1000}(a) we show the density 
profile of a system with 1000 sites, $N_f=840$, and a trapping 
potential of the form $V_{10}x_i^{10}$ with 
$V_{10}a^{10}=7 \times 10^{-27}t$. It can be seen that the density 
is almost flat all over the trap with a density of the order 
of one particle per site. Only a small part of the system
at the borders has the variation of the density required for the
particles to be trapped. In Fig.\ \ref{PerfilK1000}(b) (continuous line), 
we show the corresponding normalized momentum distribution. It 
can be seen that a kind of Fermi surface develops in the system but 
for smaller values of $k$, $n_k$ is always smooth and its value starts 
decreasing at $k=0$.
In order to see how $n_k$ changes when an incompressible region
appears in the system, we introduced an additional alternating
potential, so that in this case the new 
Hamiltonian has the form
\begin{equation}
H  = -t \sum_{i,\sigma} ( c^\dagger_{i\sigma} c^{}_{i+1
\sigma} + \textrm{H.c.} )  + V_{10} \sum_{i \sigma} x_i^{10}\ n_{i \sigma} 
+ V_a  \sum_{i \sigma} \left(-1\right)^i 
n_{i \sigma} , \label{HubbIon}
\end{equation}
where $V_a$ is the strength of the alternating potential. For the 
parameters presented in Fig.\ \ref{PerfilK1000}(a), we
obtain that a small value of $V_a$ ($V_a=0.1t$) generates a band insulator 
in the trap, which extends over the region with $n \sim 1$ (when $V_a=0$). 
However, the formation of this band insulator is barely reflected 
in $n_k$, as can be seen in Fig.\ \ref{PerfilK1000}(b) (dashed line). 
Only when the value of $V_a$ is increased and the system departs from the 
phase transition [$V_a=0.5t$, dotted line in Fig.\ \ref{PerfilK1000}(c)], 
does a quantitatively appreciable change in $n_k$ appear.

\section{Phase Diagram \label{FD}}

\begin{figure}[h]
\begin{center}
\includegraphics[width=0.6\textwidth,height=0.46\textwidth]
{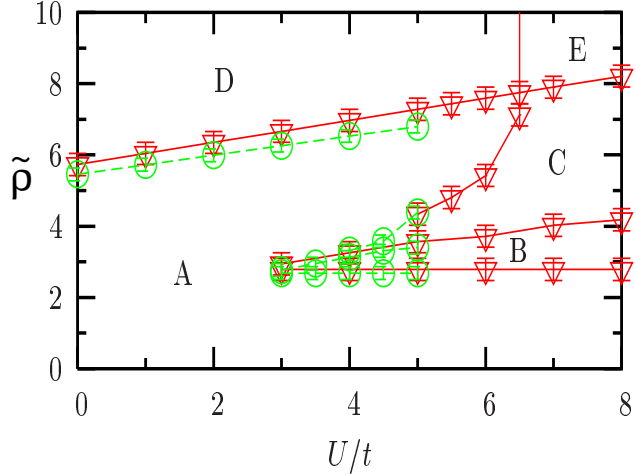}
\end{center}
\caption{Phase diagram for a system with $N=100$, $V_2a^2=0.006 t$
(\textcolor{red}{$\bigtriangledown$}) and $N = 150$, $V_2a^2=0.002 t$
(\textcolor{green}{$\bigcirc$}) sites. The phases are explained 
in the text.}
\label{PhaseD}
\end{figure}

\begin{figure}[h]
\begin{center}
\includegraphics[width=1.0\textwidth,height=0.44\textwidth]
{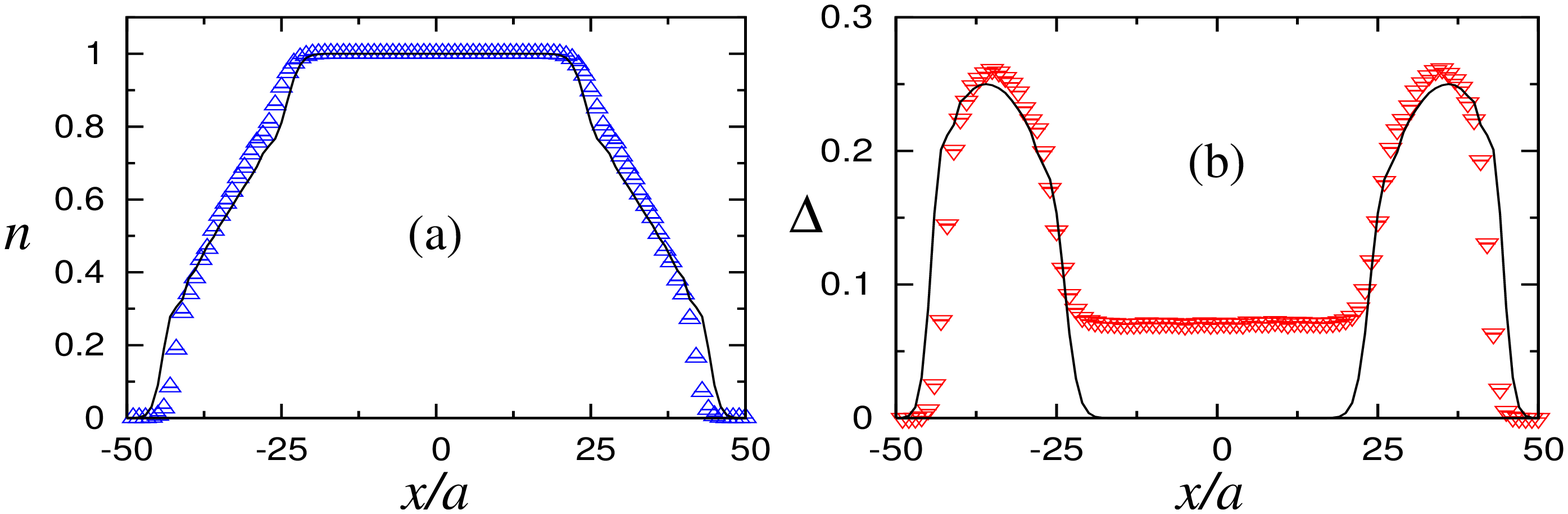}
\end{center}
\caption{Comparison between density (\textcolor{blue}{$\triangle$}) and
variance (\textcolor{red}{$\bigtriangledown$}) profiles for a system with 
$U/t=8$ and a system with $U/t=\infty$ (continuous line). In both cases 
$V_2a^2=0.0025t$ and $N_f=70$.}
\label{perfdeltInf}
\end{figure}

Finally we consider the phase diagram of the system. As shown in Figs. 
\ref{perfil3D} and \ref{perfdeltcomp100}, unlike the exponents,
phase boundaries seem to be rather sensitive to the choice of 
potential, number of particles and strength of the interaction. As in
the unconfined case, we would expect to be able to relate systems with
different number of particles and/or sizes by their density. Given the
harmonic potential, a characteristic length (in units of the lattice
constant) is given by $\zeta=\left(V_{2}/t \right)^{-1/2}$, such that a
characteristic density can be defined. Figure \ref{PhaseD} shows that
the characteristic density $\tilde{\rho}=N_f/\zeta$ is a meaningful 
quantity to characterize the phase diagram.
There, the phase diagrams for two
systems with different sizes ($N = 100$ and $N = 150$) and different
strength of the harmonic potential ($V_2a^2=0.006 t$ and $V_2a^2=0.002 t$
respectively) are depicted showing that such a scaling allows to
compare systems with different sizes, different number of particles,
and different strength of the potential. With this, it is possible to relate
the results of numerical simulations to much larger experimental
systems. The different phases obtained are: A pure metal without
insulating regions (A), a Mott-insulator at the center of the trap
(B), a metallic intrusion at the center of a Mott-insulator (C), a
``band insulator'' (i.e. with $n=2$) at the center of the trap
surrounded by a metal (D), and finally a ``band insulator'' surrounded
by a metal, surrounded by a Mott-insulator with the outermost region
being again a metal (E). Two features are remarkable here. The first 
one is that on varying the filling of the trap, a reentrant behavior is
observed for the phase A. The density profile shows a shoulder as can
be seen in Fig. \ref{perfil3D} before reaching the plateau with $n=2$
but, as shown by the inset of Fig. \ref{perfdeltcomp100} for $U=4$ around 
$n=1$, it is possible to go through a region with $n=1$ without reaching 
the value of the variance that corresponds to a Mott-insulator. The second 
intriguing feature is that the boundary between the regions A and B
remains at the same value of the characteristic density for all values
of $U$ that could be simulated. The latter feature is consistent with the 
fact that for $U\rightarrow \infty$ density properties of two-component 
fermionic systems become identical to the ones of single species fermions 
with the same total filling. In this case, the characteristic density for 
the formation of the insulating state with $n=1$ in the middle of the trap 
was obtained in Ref.\ \cite{rigol03_3} ${\tilde \rho}\sim 2.6-2.7$. This 
is the same value observed in Fig.\ \ref{PhaseD} for the formation of the 
Mott insulator. A comparison between density profiles for systems with a 
Mott insulator in the middle of the trap for $U/t=8$ and $U/t=\infty$ is 
presented in Fig.\ \ref{perfdeltInf}(a). No big differences are observed 
between both cases. This is clearly in contrast to the comparison of the 
variance profiles [Fig.\ \ref{perfdeltInf}(b)] where in the Mott plateau 
the variance has changed abruptly.

\section{Conclusions \label{Conc}}

We have studied ground state properties of two-species fermions 
confined on 1D optical lattices, reviewing and enhancing the 
analysis in Refs.\ \cite{rigol03_1,rigol03_2}. 
We have defined a local-order parameter, that we called local compressibility, 
which characterizes the local insulating regions in an unambiguous way. It
vanishes in the insulating phases and is finite in the metallic ones.
This local compressibility gives a measure of the local change in density 
due to a constant shift of the potential over finite distances larger than 
the correlation length of the density-density correlation function
for the Mott-insulating phase in the unconfined system. We found that
the local compressibility exhibits critical behavior 
(a power-law decay $\kappa^l\sim |1-n|^{\varpi}$) on entering the 
Mott-insulating regions. Due to the microscopic nature of the phases, 
spatial correlations appear not to contribute to the critical behavior 
discussed there. This is a new form of metal--Mott-insulator transition not 
observed so far in simple periodic systems, and that might be realized 
in fermionic gases trapped on optical lattices. The exponent of the 
power-law ($\varpi$) was obtained to be independent of the confining
potential and/or strength of the interaction, excluding, however, 
the unconfined case. Universal behavior was also observed for the variance 
$\Delta\sim |1-n| $ when $n \rightarrow 1$. In this case, the observed 
behavior is shared by the unconfined model. 

As opposed to periodic systems, where the appearance of the gap in the 
insulating phase can be seen by the disappearance of the Fermi surface 
in the momentum distribution function, we found that due to the presence 
of a harmonic confining potential such a feature is much less evident.
In a non-interacting case, we have shown that  
although increasing the power of the confining potential sharpens the 
features connected with the Fermi edge, it remains qualitatively different 
from the homogeneous case. Hence, in fermionic systems the momentum 
distribution function does not seem to be the appropriate quantity to 
detect experimentally the formation of Mott-insulating domains.

Finally, we determined a generic form for the phase diagram that
allows to compare systems with different values of all the
parameters involved in the model. It can be used to predict the phases that 
will be present in future experimental results. The phase diagram also 
reveals interesting features such as reentrant behavior
in some phases when parameters are changed, and phase boundaries 
with linear forms. Results obtained for finite values of the on-site 
repulsion were contrasted with the ones for infinite $U$. We observe 
that once a Mott plateau has appeared in the middle of the trap,  
density profiles do not change much by increasing the on-site repulsion. 
This is in contrast to the variance of the density, which reduces 
continuously to zero in the Mott-insulating phase.

\section*{Acknowledgments}

We gratefully acknowledge financial support from the LFSP Nanomaterialien, 
and SFB 382. We are grateful to G. G. Batrouni and R. T. Scalettar
for interesting discussions at the early stages of this project. 
We are indebted to M. Arikawa, F. G\"ohmann, and
A. Schadschneider for instructive discussions on Bethe-Ansatz.
We thank B. A. Berg for sending us a copy of Ref.\ \cite{berg92} 
on double jackknife bias-corrected estimators, which was the method we used 
to analyze our data and produce the results shown in Sec.\ \ref{PRL}.
We thank HLR-Stuttgart (Project DynMet) for allocation of computer time. 
The calculations were carried out on the HITACHI SR8000.


\begin{thebibliography}{10}

\bibitem{dalfovo99}
F. Dalfovo, S. Giorgini, L.~P. Pitaevskii, and S. Stringari, Rev. Mod. Phys.
  {\bf 71},  463  (1999).

\bibitem{leggett01}
A.~J. Leggett, Rev. Mod. Phys. {\bf 73},  307  (2001).

\bibitem{pethick02}
C.~J. Pethick and H. Smith, {\em Bose-Einstein Condensation in Dilute Gases}
  (Cambridge University Press, Cambridge, 2002).

\bibitem{pitaevskii03}
L.~P. Pitaevskii and S. Stringari, {\em Bose-Einstein Condensation} (Oxford
  University Press, Oxford, 2003).

\bibitem{anderson95}
M.~H. Anderson, J.~R. Ensher, M.~R. Matthews, C.~E. Wieman, and E.~A. Cornell,
  Science {\bf 269},  198  (1995).

\bibitem{bradley95}
C.~C. Bradley, C.~A. Sackett, J.~J. Tollett, and R.~G. Hulet, Phys. Rev. Lett.
  {\bf 75},  1687  (1995).

\bibitem{davis95}
K.~B. Davis, M.-O. Mewes, M.~R. Andrews, N.~J. van Druten, D.~S. Durfee, D.~M.
  Kurn, and W. Ketterle, Phys. Rev. Lett. {\bf 75},  3969  (1995).

\bibitem{greiner02}
M. Greiner, O. Mandel, T. Esslinger, T.~W. H\"ansch, and I. Bloch, Nature {\bf
  415},  39  (2002).

\bibitem{stoferle04}
T. St\"oferle, H. Moritz, C. Schori, M. K\"ohl, and T. Esslinger, Phys. Rev.
  Lett. {\bf 92},  130403  (2004).

\bibitem{jaksch98}
D. Jaksch, C. Bruder, J.~I. Cirac, C.~W. Gardiner, and P. Zoller, Phys. Rev.
  Lett. {\bf 81},  3108  (1998).

\bibitem{batrouni02}
G.~G. Batrouni, V. Rousseau, R.~T. Scalettar, M. Rigol, A. Muramatsu, P.~J.~H.
  Denteneer, and M. Troyer, Phys. Rev. Lett. {\bf 89},  117203  (2002).

\bibitem{kashurnikov02}
V.~A. Kashurnikov, N.~V. Prokof'ev, and B.~V. Svistunov, Phys. Rev. A {\bf 66},
   031601(R)  (2002).

\bibitem{wessel04}
S. Wessel, F. Alet, M. Troyer, and G.~G. Batrouni, cond-mat/0404552.

\bibitem{rigol03_1}
M. Rigol, A. Muramatsu, G.~G. Batrouni, and R.~T. Scalettar, Phys. Rev. Lett.
  {\bf 91},  130403  (2003).

\bibitem{rigol03_2}
M. Rigol and A. Muramatsu, Phys. Rev. A {\bf 69},  053612  (2004).

\bibitem{modugno03}
G. Modugno, F. Ferlaino, R. Heidemann, G. Roati, and M. Inguscio, Phys. Rev. A
  {\bf 68},  011601(R)  (2003).

\bibitem{ott04}
H. Ott, E. de~Mirandes, F. Ferlaino, G. Roati, G. Modugno, and M. Inguscio,
  Phys. Rev. Lett. {\bf 92},  160601  (2004).

\bibitem{anderson59}
P.~W. Anderson, Phys. Rev. {\bf 115},  2  (1959).

\bibitem{hubbard63}
J. Hubbard, Proc. R. Soc. London A {\bf 276},  238  (1963).

\bibitem{imada98}
M. Imada, A. Fujimori, and Y. Tokura, Rev. Mod. Phys. {\bf 70},  1039  (1998).

\bibitem{loh92}
E.~Y. Loh and J.~E. Gubernatis,  in {\em Modern Problems in Condensed Matter
  Sciences}, edited by W. Hanke and Y.~V. Kopaev (North-Holland, Amsterdam,
  1992), Vol.~32, pp.\ 177--235.

\bibitem{muramatsu99}
A. Muramatsu,  in {\em Quantum Monte Carlo Methods in Physics and Chemistry},
  edited by M.~P. Nightingale and C.~J. Umrigar (NATO Science Series, Kluwer
  Academic Press, Dordrecht, 1999), pp.\ 343--373.

\bibitem{assaad02}
F.~F. Assaad,  in {\em Quantum Simulations of Complex Many-Body Systems: From
  Theory to Algorithms}, edited by J. Grotendorst, D. Marx, and A. Muramatsu
  (John von Neumann Institute for Computing (NIC) Series, Vol. 10, FZ-J\"ulich,
  2002), pp.\ 99--155.

\bibitem{vignolo00}
P. Vignolo, A. Minguzzi, and M.~P. Tosi, Phys. Rev. Lett. {\bf 85},  2850
  (2000).

\bibitem{butts97}
D.~A. Butts and D.~S. Rokhsar, Phys. Rev. A {\bf 55},  4346  (1997).

\bibitem{yuval70}
G. Yuval and P. Anderson, Phys. Rev. B {\bf 1},  1522  (1970).

\bibitem{fisher89}
M.~P.~A. Fisher, P.~B. Weichman, G. Grinstein, and D.~S. Fisher, Phys. Rev. B
  {\bf 40},  546  (1989).

\bibitem{batrouni90}
G.~G. Batrouni, R.~T. Scalettar, and G.~T. Zimanyi, Phys. Rev. Lett. {\bf 65},
  1765  (1990).

\bibitem{lieb68}
E.~H. Lieb and F.~Y. Wu, Phys. Rev. Lett. {\bf 20},  1445  (1968).

\bibitem{schadschneider91}
A. Schadschneider and J. Zittartz, Z. Phys. B {\bf 82},  387  (1991).

\bibitem{rigol03_3} M. Rigol and A. Muramatsu, Phys. Rev. A {\bf 70}, 
043627 (2004). 

\bibitem{berg92}
B. A. Berg, Comp. Phys. Commun. {\bf 69}, 7 (1992).

\end{thebibliography}
\end{document}